\documentclass[10pt]{iopart}

\usepackage{amsfonts}

\usepackage{mathptmx}

\usepackage{graphicx}
\graphicspath{ {figs/} }

\usepackage{hyperref}

\usepackage{subfigure}

\newcommand{\rd}{\textrm{d}\,}

\renewcommand{\P}{\mathbb{P}} 

\newcommand{\defeq}{:=}

\newcommand{\condprob}[2]{\P\left(\left.#1 \right| #2\right)}

\newcommand{\pd}[2]{\left( \frac{\partial #1}{\partial #2} \right)}

\newcommand{\lp}[1]{l^{+}_{#1}}
\newcommand{\lm}[1]{l^{-}_{#1}}
\renewcommand{\sp}[1]{s^{+}_{#1}}
\newcommand{\sm}[1]{s^{-}_{#1}}
\newcommand{\ee}[1]{e_{#1}}

\newcommand{\C}[2]{C_{#1,#2}}
\renewcommand{\k}[1]{\kappa_{#1}}
\newcommand{\g}[2]{g_{#1,#2}}

\newcommand{\mean}[1]{\left\langle #1 \right\rangle}
\newcommand{\size}[1]{\left| #1 \right|}

\newcommand{\n}{\mathbf{n}}
\newcommand{\E}{\mathbf{E}}

\renewcommand{\L}{\mathcal{L}}

\newcommand{\Deltai}{\Delta_{(i)}}

\newcommand{\Deltaj}{\Delta_{(j)}}

\newcommand{\w}{\omega}

\newcommand{\source}{\mu}

\newcommand{\eqref}[1]{(\ref{#1})}

\newcommand{\binom}[2]{\left( {\begin{array}{c} #1 \\ #2 \\ \end{array}} \right)}

\newcommand{\text}[1]{\mathrm{#1}}

\eqnobysec

\date{\today}

\begin{document}

\title[Long-range correlations in coupled
transport]{Long-range correlations in a simple stochastic model of coupled
transport}

\author{Hern\'an Larralde$^1$ and David P.~Sanders$^2$}
\address{$^1$ Instituto de Ciencias F\'isicas, Universidad Nacional Aut\'onoma de M\'exico,
Apartado Postal 48-3, 62551 Cuernavaca, Morelos, Mexico}
\address{$^2$ Departamento de F\'isica, Facultad de Ciencias, Universidad 
Nacional Aut\'onoma de M\'exico, Ciudad Universitaria, 04510 M\'exico D.F., Mexico}
\ead{hernan@fis.unam.mx \textrm{and} dps@fciencias.unam.mx}
\date{\today}


\begin{abstract}

We study coupled transport in the nonequilibrium stationary state of a
model consisting of independent random walkers, moving along a
one-dimensional channel, which carry a conserved energy-like quantity,
with density and temperature gradients imposed by reservoirs at the
ends of the channel. In our model, walkers interact with other walkers
at the same site by sharing energy at each time step, but the amount of energy carried
does not affect the  motion of the walkers. We find that
already in this simple model long-range correlations arise in the
nonequilibrium stationary state which are similar to those observed in
more realistic models of coupled transport. 
We derive an analytical expression
for the source of these correlations, which we use to obtain
semi-analytical
results for the correlations themselves assuming a local-equilibrium hypothesis.
These are in very good agreement
with results from direct numerical simulations.
\end{abstract}
	

\section{Introduction}

Two related outstanding problems at the heart of nonequilibrium
statistical mechanics are the structure of the 
probability distribution function in the stationary state, and the derivation of macroscopic
transport laws, such as Fourier's law of heat conduction, from
microscopic dynamics, for systems which are maintained out of
equilibrium by the imposition of thermodynamic fluxes
\cite{BonettoLebowitzReyBelletFourierLawChallengeToTheorists2000,
  LepriLiviPolitiThermalConductionLowDimLatticesPhysRep2003, KomatsuExpressionStationaryDistnNoneqSteadyStatesPRL2008}.

For certain classes of stochastic mass-transport equations, known as
zero-range processes, in which the dynamics of mass leaving a site
depends only on the occupation number at that site, the stationary-state
distribution is known to factorise into the product of single-site distributions under certain conditions,
which enables many analytical results to be obtained \cite{EvansMajumdarZiaFactorizedSteadyStatesMassTransportJPA2004}.  However, for
more complicated models, this distribution no longer
factorises. In this case, the appropriate characterisation of the
stationary-state distribution becomes a central goal for the description
of these systems. Furthermore, the fact that the distribution does not
factorise implies the existence of \emph{spatial correlations} between
different sites, as has been discussed in many previous works
\cite{DorfmanSengersGenericLongRangeCorrelationsReview,
SengersHydrodynamicFluctuationsBook}.  These spatial correlations in
nonequilibrium states have been studied at a mescoscopic level using
fluctuating hydrodynamics
\cite{SengersHydrodynamicFluctuationsBook}. Generically, they tend to
be \emph{long-range}, spanning the whole length of the system, rather
than decaying exponentially as in equilibrium systems away from
critical points.

From a microscopic point of view, the correlations arising in
nonequilibrium stationary states have already been studied in many
simple models, including a stochastic master equation describing heat
flow \cite{NicolisMalekMansourOnsetSpatialCorrelationsPRA1984},
oscillators which exchange energy
\cite{KipnisMarchioroHeatFlowExactlySolvableJSP1982}, lattice-gases with
exclusion \cite{SpohnLongRangeCorrelationsStochLatticeGasJPA1983}
\cite{GarridoLebowitzLongRangeCorrelationsPRA1990}, and lattice-gas cellular
automata \cite{BoonSuarezLongRangeCorrelationsLatticeGasPRE1996}.
Also, exact results for all correlation functions were found using matrix technique for the symmetric
simple exclusion process: see e.g.\
\cite{DerridaNoneqSteadyStatesFluctnsLargeDevnsReviewJSM2007} for a
recent review, where the relation of long-range correlations to a
non-local free energy functional is also discussed.  An approximation
of the invariant measure using Gaussians in a suitably rotated
coordinate system has recently been obtained
\cite{DelfiniLiviPolitiNoneqInvarMeasureHeatFlowPRL2008,LepriMejiaPolitiStochModelAnomHeatTransportJPA2009}; and related
analytical results were previously found by Su\'arez et al.\
\cite{BoonSuarezLongRangeCorrelationsLatticeGasPRE1996}, for the case
where transport is by particles with exclusion, but with a single
conserved quantity.

In particular, long-range correlations in the so-called
``random-halves'' model \cite{EckmannYoungNoneqProfilesCMP2006} of
\emph{coupled} matter and heat transport were recently studied in
\cite{LinYoungCorrelationsRandomHalvesJSP2007}, principally
heuristically and numerically. The model we study in this paper can be
considered to be a simplified version of the random-halves model,
still containing two explicitly conserved quantities.  The
simplification enables us to obtain an explicit expression for the
source of the correlations in the nonequilibrium stationary state of the system.

Although our model suppresses much of the physical meaning of the
second conserved quantity, in addition to mass, which in
\cite{LinYoungCorrelationsRandomHalvesJSP2007} can really be viewed as
corresponding to energy, we emphasise that the structure of the
spatial correlations that we observe for this energy-like quantity is
remarkably similar to that found in
\cite{LinYoungCorrelationsRandomHalvesJSP2007}.

In this paper, we study the equilibrium and nonequilibrium stationary
states of the model. In regards to the nonequilibrium stationary state, we
obtain the transport equations for the energy and mass, and we obtain
the equation satisfied by the spatial energy correlations that arise
in the model. This equation has a non-closed form.  To close the
hierarchy, we make a local equilibrium assumption, which enables an
analytical evaluation of the form of the correlation source terms. We
are then left with an approximate discrete Poisson equation with
source terms for the correlations. We find very good agreement between
the solution of this equation with numerical simulations of the
system.


\section{Coupled transport model}

In this section we introduce the model of transport which we
study. It is, perhaps, one of the simplest stochastic models
exhibiting coupled transport.  The transported quantities are
particles (mass), and a second quantity, which is locally conserved,
which the particles carry with them when they move.  For simplicity of
exposition, we refer to this second quantity as ``energy'', although
we emphasise that it does not necesarily have the physical
characteristics of an energy, since the motion of the walkers is
\emph{independent} of the value of the energy which they carry.

Specifically, the model consists of independent random walkers moving
on a one-dimensional chain of $L$ sites.  The system is open, and is in contact
with particle and energy baths at each end of the chain,
which at each time
step supply or remove particles from the system with a given rate and energy
distribution corresponding to their density and temperature, respectively.

The walkers move synchronously in discrete time: at each time step,
each walker independently attempts to jump to one of its two
neighbouring sites, or remains at the same site.
If a walker successfully jumps, then it carries with it an
amount of energy $s$ from the total amount of energy $E$ at its
previous site.

After all particles have attempted their jumps, the total energy at
each site, that is, the sum of the individual energies of the
walkers at that site, is redistributed among all the particles at that
site, in a ``random'' (microcanonical) way, which we specify below.
We thus have a complete (infinite) separation of time-scales: energy
equilibration at each site is completed before the particles move
again.  This separation of time scales is, in part, what enables us to
proceed with the analysis of the system. Further, it ensures that we
can always use a local equilibrium hypothesis, in the sense that all
thermodynamic quantities are always well defined at every site and
that they are related to each other according to the usual
(equilibrium) thermodynamic relations.

As mentioned above, our model is related to the recently-introduced
\emph{random-halves} model \cite{EckmannYoungNoneqProfilesCMP2006,
LinYoungCorrelationsRandomHalvesJSP2007}, designed to model, rather
faithfully, the Hamiltonian dynamics underlying the transport
phenomena observed in
\cite{MejiaLarraldeLeyvrazCoupledTransportPRL2001,
LarraldeLeyvrazMejiaTranspPropsModifiedLorentzJSP2003}.  In the
random-halves model, each particle jumps to a neighbouring site with a
rate which is proportional to the square root of its kinetic energy,
and which is a factor $\delta$ times the rate at which a particle
exchanges a random fraction of its energy with a reservoir located at
its current position.

Taking the limit $\delta \to 0$ in the random-halves model corresponds
to our limit of infinite separation of time-scales, the other
particles at that site acting as the reservoir.  However, the
random-halves model includes extra effects which our model cannot
account for: for example, in the random-halves model, it is possible
to have sites with particles at very low energy, and since the jump
rate is energy dependent, these particles may remain a long time at
that site unless other very energetic particles arrive there.
Nonetheless, as we shall see, these kind of effects do not appear to
affect the qualitative results on correlations.

Although we do not consider it in this work, it should be noted that if
we make the jump probabilities $p$ and $q$ very small (of order
$1/N$, where $N$ represents the number of particles in the system) we
effectively recover single-particle motion, as in continuous-time dynamics,
that is, on average, only one particle moves at each time step.
Furthermore, in this case we could unambiguously consider
jumping probabilities that are functions of the energy of the single
moving particle, which could yield a model closer to that considered
in \cite{LinYoungCorrelationsRandomHalvesJSP2007}. However, such
modifications render the system intractable, and do not appear to be a
necessary ingredient for the presence of long-range correlations in the
nonequilibrium stationary states.

\subsection{Master equation}




We now proceed to specify the model precisely.
We consider an arbitrary number of random walkers which can occupy
sites on a finite one-dimensional chain of sites, labelled by $i \in
\{1,\ldots,L\}$. 
 The system is open, and is in contact with particle and energy
baths at sites $0$ and $L+1$. The baths have mean particle densities
$\rho_0$ and $\rho_{L+1}$, and are at temperatures $T_0$ and
$T_{L+1}$. This means that the number of particles available in each
bath is drawn from a Poisson distribution with mean $\rho_0$ and
$\rho_{L+1}$, respectively, at each time step, and the energy $E$
carried by each particle leaving a bath at temperature $T$ has a
Boltzmann distribution at that temperature, $P(E) = \frac{1}{T}
e^{-E/T}$.

 Let $n_i$ and
$E_i$ be the
number of particles and the total energy at site $i$ at time $t+1$,
and $m_i$ and $e_i$ the corresponding quantities at time $t$.  The
walkers can jump to the right with probability $p$, jump to the left
with probability $q$, or remain where they are with probability
$r\defeq 1-(p+q)$.  The number of walkers which jump right from site
$i$ at a given time step is a random variable denoted $\lp{i}$, and
similarly, the number jumping left from that site is $\lm{i}$.  All
walkers jump simultaneously.


Each walker carries a certain amount of energy with it when it jumps.  After
each step, the new total energy $E_i$ at a site $i$ is distributed randomly
among the $n_i$ walkers at that site, according to a ``microcanonical
distribution''.  The total amount of energy carried by the walkers which move
from site $i$ to the right is denoted by $\sp{i}$, and to the left by
$\sm{i}$.

The master equation describing the time evolution of this system is
then given by

\begin{eqnarray}
\eqalign{
\fl P_{t+1}(n_1,E_1; n_2,E_2; \ldots; n_L,E_L) = \\
\sum_{\{m_i\}} \sum_{\{l_i^{\pm}\}} \int_{\{e_i\}} \rd e_i \int_{\{s_i^{\pm}\}}
\rd s_i^{\pm} \,
P_t(m_1, e_1; m_2,e_2; \ldots; m_L,e_L) \\
\times \prod_i \delta\left(n_i-[m_i + (\lp{i-1} + \lm{i+1}) - (\lp{i} +
\lm{i})]\right) \\
\times \prod_i \delta \left(E_i-[e_i + (\sp{i-1}+\sm{i+1}) - (\sp{i}+\sm{i})]
\right) \\
\times \prod_i \condprob{\sp{i},\sm{i}} {\lp{i},\lm{i},m_i, e_i} 
\times \prod_i \condprob{\lp{i},\lm{i}} {m_i}.
}
\label{master}
\end{eqnarray}
The delta functions reflect the fact that the new occupation numbers
and energies are obtained from the old ones by the movements at that
time step.  The conditional probabilities appearing in the last line of
this equation denote the probability densities for the number
of walkers and total energy moving left and right, and are given by
\begin{eqnarray}
\condprob{\lp{i},\lm{i}} {m_i} \defeq
	\binom{m_i}{\lp{i}} \binom{m_i-\lp{i}}{\lm{i}} \, p^{\lp{i}} q^{\lm{i}}
r^{m_i-\lp{i}-\lm{i}} \\
\eqalign{
\fl
\condprob{\sp{i},\sm{i}} {\lp{i},\lm{i},m_i, e_i} \defeq \\
\fl \frac{\Gamma(m_i)}{\Gamma(\lp{i})\Gamma(\lm{i}) \Gamma(m_i-\lp{i}-\lm{i})} 
(\sp{i})^{\lp{i}-1}  (\sm{i})^{\lm{i}-1}
\frac{(e_i-\sp{i}-\sm{i})^{m_i-\lp{i}-\lm{i}-1}}{e_i^{m_i-1}}.
}
\end{eqnarray}
The first is a trinomial distribution which gives the probability
of moving exactly $\lp{i}$ particles to the right and $\lm{i}$ to the
left, out of the $m_i$ particles at site $i$. The second
``multivariate beta distribution'' is chosen to reflect the
partitioning of the energy amongst the $\lp{i}$, $\lm{i}$ and the
remaining $m_i-\lp{i}-\lm{i}$ particles, under the assumption that
within each site the particles behave as a $2$-dimensional
ideal gas. 

The ``ideality'' of the gas at each site is manifested by the fact that the distribution can be
written exactly in terms of products of appropriate phase space
volumes, while the value of the exponents reflects the fact that the gas 
is 2-dimensional; details are given in the Appendix. This particular distribution was
chosen because it yields slightly simpler expressions (than, say, 1 or
3 dimensional ideal gases) and is closer to the intrinsic
2-dimensional nature of various previous models for coupled transport.

\subsection{Equilibrium state}\label{sec:eqm}

It is known \cite{LarraldeResendizStatsDiffusiveFlux2005} that many
non-interacting walkers, even when subjected to a density gradient,
attain a stationary state which factorises: the probability of having
occupation numbers $\n \defeq (n_1, \ldots, n_L)$ is given by the following
product of Poisson distributions at each site:
\begin{equation}
 P(\n) = \prod_{i=1}^L P(n_i), 
\end{equation}
where 
\begin{equation}
P(n_i) =  \frac{e^{-\rho_i} \rho_i^{n_i}}{n_i!},
\end{equation}
with $\rho_i$ the mean occupation number at site $i$.  The $\rho_i$
satisfy a discrete diffusion equation, which in the stationary state
becomes $\rho_i=p \rho_{i-1} + q \rho_{i+1} + r\rho_i$, and which can be
solved in terms of the boundary conditions.

Suppose now that there is no gradient of temperature imposed at the
boundaries of our model for coupled transport, i.e. $T_0 = T_{L+1} =
T$. Then it turns out the joint probability distribution of having energy
$E_i$ and $n_i$ particles at sites $i=1,2...$ is also given by a factorised
distribution:
\begin{equation}\label{fact}
P(\E; \n) = \prod_{i=1}^L P(n_i) P(E_i | n_i),
\end{equation}
where the conditional probability of have energy $E$ at a site with
$n$ particles is given by
\begin{equation} \label{eq:eqm-energy-distn}
 P(E | n) = \frac{\beta^n E^{n-1} e^{-\beta E}}{\Gamma(n)},
\end{equation}
where $\beta \defeq 1/T$. (We take units such that the Boltzmann constant $k_B=1$
throughout the paper.)  This distribution can be interpreted as 
$\Omega(E,n) e^{-\beta E}/Z(\beta,n)$, where $\Omega(E,n)$ is the volume
of phase space accesible to a 2-dimensional ideal gas of $n$
particles at total energy $E$, and $Z(\beta,n)$ is the partition
function.

The mean energy for this equilibrium distribution is $\mean{E | n} = n
/ \beta$, so that in equilibrium the mean energy at a site with mean
concentration $\rho$ is $\rho / \beta = \rho T$.  Since the
distribution of energy is that of a system with temperature $T$, we
can unambiguously identify $\beta$ with the inverse temperature.

That the distribution factorises in equilibrium can be verified by
assuming that the solution has a form as given in \eqref{fact} as an
\emph{ansatz}. It then transpires that the only way it can do so is if
the temperature profile is flat. Hence, in the \emph{presence} of a
temperature gradient the joint distribution $P(\E; \n)$ of all
energies and positions \emph{does not} factorise,
and thus spatial correlations are present.

\section{Thermodynamics}

In this section we study the thermodynamic properties of the
system. This is straightforward since, by construction, at each time
step the system reaches a microcanonical equilibrium at each site $i$,
characterized by the number of particles, $n_i$, and the energy,
$E_i$, found at that site. 

Since we are assuming that at each site the particles are a
2-dimensional ideal gas, and accounting for the indistinguishability
of the particles, the classical entropy at each site is given by
\begin{equation}
S_i = n_i \ln\left(\frac{V E_i}{n_i^2}\right) + n_i s_i, 
\label{entropy}
\end{equation}
where $V$ is the volume (actually, the area) available for the gas at
each site, which we take as unity ($V=1$), and $s_i$ is a constant, in
the sense that it is independent of the thermodynamic variables,
though it may vary from one site to another (see Appendix).

Having the fundamental relation \eqref{entropy}, we can proceed to
obtain the equations of state for the intensive variables in the entropy
representation \cite{CallenThermodynamicsBook1985}:
\begin{equation}
 \frac{1}{T_i} = \pd{S_i}{E_i}_{N_i} = \frac{n_i}{E_i}
\label{temp}
\end{equation}
and
\begin{equation}
 \frac{-\mu_i}{T_i} = \pd{S_i}{N_i}_{E_i} = \ln
 \left(\frac{E_i}{n_i^2}\right) + \nu_i, 
\label{mu}
\end{equation}
where $\nu_i \defeq s_i -2$ is a constant, which again may have
different values at different sites. These expressions will be useful
further on, in connection with the Onsager relations and the use of the
local equilibrium hypthesis.


\subsection{Concentration and energy profiles}

We now consider the case in which the system is forced out of
equilibrium by imposing concentration and/or temperature differences
at the boundaries, that is, by imposing $\rho_0 \neq \rho_{L+1}$
and/or $T_0 \neq T_{L+1}$.  If we do so, then the system will
eventually attain a nonequilibrium stationary state, with well-defined
concentration and mean energy profiles, $\rho_i$ and $\mean{E_i}$, as
a function of the position $i$ in the system. Related profiles have
been studied in detail for random-halves and other stochastic models
in \cite{EckmannYoungNoneqProfilesCMP2006,EckmannYoungTempProfilesHamiltonianHeatCondEPL2004}.

The transport equations can be easily derived by recalling that the
total energy $E_i$ at site $i$ at time $t+1$ is given by the energy $e_i$
that was there at time $t$, plus the energy brought in by the walkers
that arrived in that time step, minus the amount taken by the walkers
that left:
\begin{equation}
 E_i = e_i + (\sp{i-1} + \sm{i+1}) - (\sp{i} + \sm{i}).
\end{equation}
Taking means, we obtain
\begin{equation}
\fl \mean{E_i} = \mean{\ee{i}} + p \mean{\ee{i-1}} + q \mean{\ee{i+1}} - p
\mean{\ee{i}} - q \mean{\ee{i}} = p\mean{\ee{i-1}} + q \mean{\ee{i+1}} +
r\mean{\ee{i}}.
\end{equation}
A similar equation holds for particle transport. In the stationary state,
$ \mean{E_i} = \mean{\ee{i}}$, and hence the stationary
profiles satisfy
\begin{eqnarray}
\mean{n_i}&= p\mean{n_{i-1}} + r \mean{n_i} + q \mean{n_{i+1}};\\
\mean{E_i}&= p\mean{E_{i-1}} + r \mean{E_i} + q \mean{E_{i+1}}.
\end{eqnarray}

We denote by $\rho_i \defeq \mean{n_i}$ the stationary mean occupation
number at site $i$, and by $T_i \defeq \mean{E_i} / \rho_i$ the local
temperature there.

\subsection{Thermodynamic fluxes and forces}
\label{sec:therm-forces}

The mean energy and particle fluxes between sites $i$ and $i+1$ are given by
\numparts
\begin{eqnarray}
J_u &= p\mean{E_i}-q \mean{E_{i+1}}=p\rho_i T_i-q \rho_{i+1} T_{i+1},\\ 
J_\rho &= p\rho_i-q\rho_{i+1}.
\end{eqnarray}
\endnumparts
To obtain the continuum (diffusive) limit, we first express
$\rho_i$, $T_i$, $\rho_{i+1}$ and $ T_{i+1}$ as Taylor series around
position $x=(i+1/2) \, \delta x$, where $\delta x$ is the distance between
neighbouring sites on the chain. Next, we transform 
$\rho \to c\, \delta x$, $J_{u} \to j_{u} \, \delta t$ and $J_{\rho} \to
j_{\rho} \, \delta t$, where $\delta t$ is the
time interval between succesive steps, so the quantities $c$, 
$j_u$ and $j_\rho$ are a proper density and fluxes, respectively. These
operations yield
\numparts
\begin{eqnarray}
j_u \, \delta t &= (p - q) \, c T \, \delta x & - \textstyle \frac{1}{2} (p+q)
(\delta x)^2
\, \nabla (c T)  + \mathcal{O}(\delta x^3) \\ 
j_\rho \, \delta t &= (p - q) \, c \, \delta x & - \textstyle  \frac{1}{2} (p+q)
(\delta x)^2
\, \nabla c  + \mathcal{O}(\delta x^3).
\end{eqnarray}
\endnumparts

The continuum limit is achieved by dividing through by $\delta t$ and
taking the limit in which $\delta t$, $\delta x$ and  $p - q$ tend to $0$, in
such a way
that the ratios $(\delta x)^2/\delta t$ and $(p - q)/\delta x$ remain
finite.  Thus, we can define the drift velocity 
\begin{equation}
v \defeq \lim_{\delta t \to 0,\, \delta x \to 0}(p-q) \, \delta x/\delta t
\label{drift}
\end{equation}
and the diffusion constant 
\begin{equation}
D \defeq \textstyle   \frac{1}{2} \displaystyle \lim_{\delta t \to 0,\, \delta x
\to 0} (p+q) \,
(\delta x)^2/\delta t,
\label{diff}
\end{equation}
in terms of which the above
equations become
\numparts
\begin{eqnarray}
j_u &= - D \, \nabla (c T)+ v c T
\label{eq:first-set-1} 
\\
j_\rho &= - D \, \nabla c + v c.
\label{eq:first-set-2}
\end{eqnarray}
\endnumparts
These should be compared with
\numparts
\begin{eqnarray}
j_u &= L_{11} \nabla (1/T)+ L_{12}\nabla(-\mu/T)\\
j_\rho &= L_{21} \nabla (1/T)+ L_{22}\nabla(-\mu/T)
\end{eqnarray}
\endnumparts
from the theory of linear thermodynamics
\cite{CallenThermodynamicsBook1985, DeGrootMazurNonEquilibriumThermodynamics1984}. 
Using \eqref{temp} and \eqref{mu}, we obtain
\numparts
\begin{eqnarray}
j_u &= -L_{11}(\nabla T)/T^2 
+ L_{12}\left[(\nabla T)/T -(\nabla c)/c +\nabla \nu\right]
\label{eq:second-set-1}
\\
j_\rho &= -L_{21}(\nabla T)/T^2 
+ L_{22}\left[(\nabla T)/T -(\nabla c)/c +\nabla \nu\right].
\label{eq:second-set-2}
\end{eqnarray}
\endnumparts
%
 Setting $\nabla T=\nabla c=0$, we find
\begin{equation}
L_{12}\nabla \nu =v c T \qquad{\rm and}\qquad
L_{22}\nabla \nu= v c.
\end{equation}
If instead we set $\nabla T=\nabla\nu=0$, then 
\begin{equation}
L_{12}/ c= D T \qquad \mathrm{and} \qquad  L_{22}/c=D.
\end{equation}
Finally, if we set $\nabla c=\nabla\nu=0$, then
\begin{equation}
L_{11}/T^2-L_{12}/T = D c \qquad \mathrm{and} \qquad  L_{22}/T-L_{21}/T^2=0.
\end{equation}
From these equations, we obtain:
\begin{eqnarray}
L_{11}= 2 D T^2 c; &\qquad L_{12}=L_{21}= D T c; \\
L_{22}= D c;   &\qquad \nabla \nu= v/D.
\end{eqnarray}
Thus the Onsager reciprocal relations
\cite{DeGrootMazurNonEquilibriumThermodynamics1984}
are satisfied, and $\nu_i$ is determined as
an external potential due to the overall current generated by the bias.

While these results are satisfactory, it should be noted that we
made a rather cavalier use of \eqref{temp} and \eqref{mu}, namely,
we identified the temperature at site $i$ as the quantity
$\mean{E_i}/\mean{n_i}$, whereas \eqref{temp} tells us that the local
temperature is actually the stochastic variable $T_i=E_i/n_i$;
furthermore, we substituted the remaining $n_i$ in \eqref{mu} by
$c_i \, \delta x=\rho_i=\mean{n_i}$. These substitutions are, of course, not
generally
valid; however, they can be justified when the fluctuations in energy
and number of particles are small compared to their mean values. 

\section{Spatial correlations of the energy}

We now turn to the main consideration of the paper, the origin of the spatial correlations between the
values of energy at different sites, which develop due to the
imposition of a temperature gradient. To this end,
we denote by $\C{i}{j}\defeq \mean{E_i E_j} - \mean{E_i}\mean{E_j}$
the stationary-state energy correlations between sites $i$ and $j$.  
To simplify the notation, we use the difference operators
$\Deltai$ and $\Deltaj$ which act on functions of two
variables $\C{i}{j}$ as
\begin{eqnarray}
[\Deltai C ] _{i,j}  \defeq p \C{i-1}{j} + r \C{i}{j} + q \C{i+1}{j}\ ; \\[0pt]
[\Deltaj C ] _{i,j}  \defeq p \C{i}{j-1} + r \C{i}{j} + q \C{i}{j+1}\ .
\end{eqnarray}

\subsection{Exact equation for stationary-state energy correlations}

Using the above notation, it follows from the previous section that
the evolution equation for the average energy is $\mean{E_i} = \Deltai
\mean{e_i}$, and thus
\begin{equation}
\mean{E_i} \mean{E_j} = \Deltai \Deltaj \left[\mean{e_i}
\mean{e_j} \right],
\end{equation}
so that this part of the correlations $\C{i}{j}$ factorises.

It remains to evaluate $\mean{E_i E_j}$.  To do so, we rewrite it
quantity in terms of the energies $\ee{i}$ and $\ee{j}$ at sites $i$ and
$j$ before the move, and the amounts of energy moving in each direction
from each site:
\begin{equation}
\fl \mean{E_i E_j} = \mean{\left[ \ee{i} + (\sp{i-1} + \sm{i+1}) - (\sp{i} +
\sm{i}) \right].
\left[  \ee{j} + (\sp{j-1} + \sm{j+1}) - (\sp{j} +
\sm{j}) \right]
}.
\end{equation}
We expand the product and consider the resulting terms, which are
means of products of two random variables, of the form
$\mean{\sp{i-1}\sp{j+1}}$.  According to the master equation
\eqref{master}, these random variables are independent if their indices
are different, giving, for example, $\mean{\sp{i-1}\sp{j+1}} =
\mean{\sp{i-1}} \mean{\sp{j+1}}$ if $i-1 \neq j+1$.  In particular,
this is the case for every pair of products provided $\size{i-j} > 2$.

If, on the other hand, $|i-j| \le 2$, then there are terms in the
product for which the indices are the same: for example, if $j=i+1$,
then $\sm{i+1} = \sm{j}$.  In this case, the mean of the product is no
longer the product of the means, and we must calculate it
explicitly.  For example, we have
\begin{equation}
 \mean{\sp{i} | e_i, \lp{i}, m_i } = \frac{\lp{i}}{m_i} e_i; \qquad
\mean{\sp{i}{}^2 | e_i, \lp{i}, m_i} = \frac{\lp{i}(\lp{i}+ 1)
e_i^2}{m_i (m_{i}+1)}.
\end{equation}
 We must then average the expressions over the trinomial
distribution for the $\lp{i}$ and $\lm{i}$ given $m$.
Note that the right-hand side of the second equation is not the square
of the first equation -- a correction term has arisen. These
corrections are what eventually give rise to the long-range energy
correlations. 

We finally obtain, after some messy algebra, which we confirmed via a
computer algebra package, the following \emph{exact} equation for the
spatial correlations $\C{i}{j}$ in the stationary state:
 \begin{equation}
\label{eq:eqn-for-C}
 \C{i}{j} = \Deltai \Deltaj \C{i}{j} + 2 \lambda_{ij}, 
\end{equation}
with $\lambda_{ij}$ a symmetric matrix given by
\begin{equation}
 \fl 
\lambda_{ij} =
\cases{
p(1-p) \k{i-1} + r(1-r) \k{i} +
q(1-q)  \k{i+1} & if $j=i$;\\
-pr \k{i} - qr \k{i+1} & if $j=i + 1$;\\
-pq  \k{i+1} & if  $j=i + 2$;\\
0 & otherwise,
}
\end{equation}
where we have defined
\begin{equation}
 \k{i} \defeq \mean{\frac{e_i^2}{m_i + 1}}.
\end{equation}
Equation \eqref{eq:eqn-for-C} is essentially a discrete Poisson equation,
with source terms $2 \lambda_{ij}$.

The boundary conditions are $\C{i}{j} = 0$ whenever $i$ or $j$ is
equal to $0$ or $L+1$, since the stochastic reservoirs at positions
$0$ and $L+1$ are independent of all other quantities in the system
(and of each other).  The exceptions to this are $\C{0}{0}$ and
$\C{L+1}{L+1}$, which are given by the variances of the distributions
in the reservoirs.

The above equations may be simplified by introducing
\begin{equation}
 \g{i}{j} \defeq \C{i}{j} - 2 \delta_{ij} \k{i}, 
\label{eq:defn-g}
\end{equation}
where $\delta_{ij}$ is the Kronecker delta, which is $1$ when $i=j$
and $0$ otherwise.  Substituting this expression in
\eqref{eq:eqn-for-C} gives that $\g{i}{j}$ satisfies the following
simpler equation:
\begin{equation}
\label{geq}
 \g{i}{j} = \Deltai \Deltaj \g{i}{j} + 2 \source_i \delta_{ij},
\end{equation}
with source terms
\begin{equation}
 \source_i \defeq p \k{i-1} + q \k{i+1} + (r-1) \k{i}.
\label{eq:defn_mu}
\end{equation}
Note that the only source terms in \eqref{geq} are now on the
diagonal.  The boundary conditions are $\g{i}{j} = 0$ when $i=0$,
$i=L+1$, $j=0$ or $j=L+1$.  We thus have a discrete Poisson equation in a square,
with zero boundary conditions and a line source term on the diagonal.

We can test this equation in the simplest case: that in which there is
no energy (temperature) gradient. In this case, the reservoirs are at
the same temperature $\beta \defeq \beta_0 = \beta_{L+1}$, so that in
fact the temperature is constant throughout the system, $\beta_i =
\beta $ for all $i$.  Under these conditions, we know that the energy
distribution factorises, hence there must be no energy
cross-correlations.  Indeed, in this case we can evaluate $\k{i}$
exactly to obtain $\k{i} = \rho_i / \beta^2$, and \eqref{eq:defn_mu}
then gives
\begin{equation}
\source_i = \frac{1}{\beta^2} \left[ p \rho_{i-1} + q \rho_{i+1} +
(r-1) \rho_{i} \right] = 0,
\end{equation}
since the $\rho_i$ satisfy precisely this discrete equation.  Thus, in
the absence of a temperature gradient, the $\g{i}{j}$ satisfy $
\g{i}{j} = \Delta_{(i)} \Delta_{(j)} \g{i}{j}$ for all $i$ and $j$,
with no source terms.  The zero boundary conditions then imply that
$\g{i}{j}$ is identically zero.

Substituting this result back into \eqref{eq:defn-g}, we obtain in
this constant temperature case
\begin{equation}
 \C{i}{j} = 2 \delta_{ij} \k{i} = 2 \frac{\rho_i}{\beta^2} \delta_{ij}.
\end{equation}
The term $2 \delta_{ij} \k{i}$ can thus be regarded as the
contribution to the energy correlation matrix which arises simply
because $\C{i}{i}$ necessarily has a non-zero on-site value, given by
\begin{equation}
\C{i}{i} = \mean{E_i^2} - \mean{E_i}^2 = 2 \k{i} = \frac{2\rho_i}{\beta^2}.
\end{equation}

Referring back to the definition \eqref{eq:defn-g} of $\g{i}{j}$, we
see that this quantity can thus be viewed as containing the long-range
part of the correlations, resulting from the imposition of temperature
gradients.  This is similar to results of previous work in the case of
a single transported quantity
\cite{NicolisMalekMansourOnsetSpatialCorrelationsPRA1984,
  BoonSuarezLongRangeCorrelationsLatticeGasPRE1996}.

We remark that the physical meaning of the terms $\k{i} = \mean{e_i^2
  / (m_i + 1)}$, which form the source terms of the long-range
correlations, and thus in some sense are what gives rise to these
correlations, is not very clear.

\subsection{Local thermodynamic equilibrium approximation}

The previous calculation is exact; however, to make further progress,
we must make an approximation in order to evaluate the terms $\k{i}$
appearing in the expression for the source $\source_i$ of the
long-range part of the correlations when the system is in a
nonequilibrium stationary state.  To do so, we will \emph{assume} that
\emph{local thermodynamic equilibrium} is attained at each site. By
this we mean the assumption that the marginal distribution of the
energy at each site $i$ is given by $P(E_i | n_i)$, with the
distribution \eqref{eq:eqm-energy-distn} which is found at
equilibrium. This is an uncontrolled approximation; however, we will
see later that it is in very good agreement with the numerical
results.  Note that $\k{i}$ involves only data at site $i$, and thus
indeed depends only on the marginal distribution at that site.  Such a
local thermodynamic equilibrium assumption has recently been proved
correct for the random-halves model, in the limit when the number of
sites goes to $\infty$, so that the temperature gradient goes to zero
\cite{RavishankarYoungLocalThermEqRandomHalvesJSP2007}.

Under the hypothesis of local thermodynamic equilibrium, we can use
\eqref{eq:eqm-energy-distn} to calculate $\k{i} = \mean{E_i^2/(m_i+1)}$,
obtaining
\begin{equation}
\k{i} =
\frac{\rho_i}{\beta_i^2} = \rho_i T_i^2. 
\label{eq:local-eqm-for-kappa}
\end{equation}
We will use this approximate analytical form for $\k{i}$ in the
remainder of the analytical development.

From the above discussion, we see that the contributions to the
onsite correlations $\C{i}{i}$ split into two parts: 
\begin{equation}
 \C{i}{i} = \g{i}{i} + 2 \k{i}.
\end{equation}
We can regard $2 \k{i}$ as a pure local contribution, and $\g{i}{i}$
as the onsite part of the long-range contribution.  Within the local
equilibrium approximation, we then obtain
\begin{equation}
 \g{i}{i} = \C{i}{i} - 2 \rho_i T_i^2.
\end{equation}

The equations for the profiles of mean concentration and mean energy
can be written as follows:
\begin{eqnarray}
 p \rho_{i-1} + q \rho_{i+1} + (r-1) \rho_{i} &= 0;\\
 p \rho_{i-1} T_{i-1} + q \rho_{i+1} T_{i+1} + (r-1) \rho_{i} T_{i} &= 0.
\end{eqnarray}
For brevity, we introduce the linear operator $\L_i [f] \defeq p f_{i-1}
+ q f_{i+1} + (r-1) f_{i} = (\Deltai - 1) f$.  The equations for the profiles
then become $\L_i[\rho] = 0$ and $\L_i [\rho T] = 0$. 

The correlation source is $\source_i = \L_i[\kappa] = \L_i[\rho T^2]$,
where the latter equality again assumes the local thermodynamic
equilibrium approximation.  Substituting $\L_i [\rho] = 0$ and
$\L_i[\rho T] = 0$ into the expression for $\L_i[\rho T^2]$, we obtain
that the source term $\source_i$ in the local thermodynamic
equilibrium approximation is given by
\begin{equation}
\fl \source_i = \L_i[\rho T^2] = \textstyle \frac{1}{2} (T_{i+1}-T_{i-1}) \left[ p
\rho_{i-1} (T_i - T_{i-1}) + q \rho_{i+1} (T_{i+1} - T_{i}) \right].
\end{equation}

In the continuum diffusion limit, we have
\begin{equation}
 \L[f] \simeq \left[D f'' - v f'\right] \, \delta t,
\end{equation}
with $v$, $D$ are defined in equations \eqref{drift} and \eqref{diff},
respectively, and $\rho= c \, \delta x$ as before. Then the source of
correlations $\source(x)$ reduces to
\begin{equation}
 \source(x) \simeq \L[ \rho T^2] \simeq 2D c (x) [T'(x)]^2 \, \delta t \, \delta
x,
 \label{eq:continuum-source}
\end{equation}
as can also be verified directly from the continuum expressions. 
We remark that precisely a quadratic dependence on the local
temperature gradient of short-range energy correlations was found
numerically for the random-halves model
\cite{LinYoungCorrelationsRandomHalvesJSP2007}.

It should be noted that only the energy has long-range correlations:
a calculation similar to the above shows that the density correlations
$\mean{n_i n_j}$ and the density--energy cross-correlations $\mean{E_i
  n_j}$ are both diagonal.

\section{Numerical results}

In this section, we present comparisons of the energy correlations as
obtained from direct simulations of the microscopic random-walk
dynamics, with the approximate analytical results derived in the
previous section.

\subsection{Numerical method}
The boundary conditions in the numerical simulations are as follows.
At each time step, the number of particles $n_0$ at the left bath is
chosen from a Poisson distribution with mean $\rho_0$, and each of
those particles is assigned an energy $E$ with probability $1/T_0
e^{-E/T_0}$.  The same is done at the right bath with appropriate
temperature and density.  At each site, the energy of the particles is
assigned via the microcanonical redistribution mechanism, and then
each particle jumps to a neighbouring site with the correct
probabilities.
Means and correlations are determined by time averaging, after
discarding a preliminary equilibration period.

The correlations from the direct numerical simulations are compared to
``semi-analytical'' results obtained by solving the discrete Poisson
equations \eqref{eq:defn-g} for the long-range part $g$ of the
correlations, using the local equilibrium approximation
\eqref{eq:local-eqm-for-kappa} for the terms $\k{i}$.  A similar
numerical solution of the algebraic equations was recently
employed in \cite{DelfiniLiviPolitiNoneqInvarMeasureHeatFlowPRL2008}.

\subsection{Temperature gradient in absence of concentration gradient}

\begin{figure}[t]
\subfigure[]{
\includegraphics{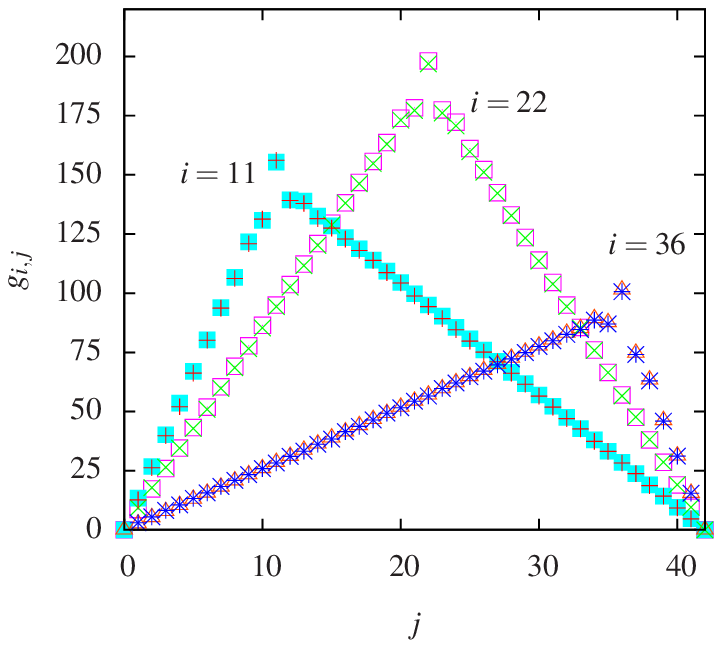}
\label{fig:temp-gradient-no-density-gradient}
}
\subfigure[]{
\includegraphics{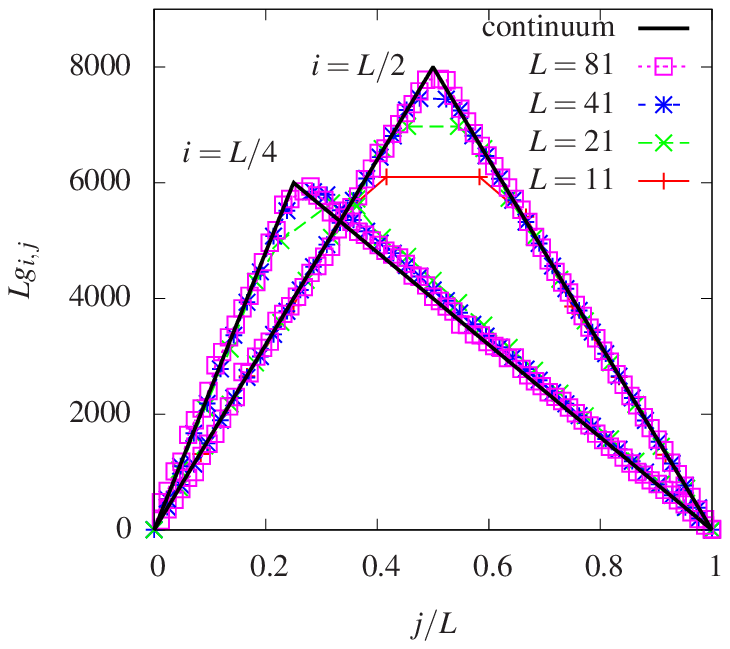}
 \label{fig:scaling}
}
 \caption{(a) Long-range part of the correlations, $\g{i}{j} = C_{i,j}
   - 2 \delta_{ij} \kappa_i$, for an imposed temperature gradient and
   flat density profile, with $L=41$, $\rho=10$, $T_0 = 50$,
   $T_{L+1}=10$. Shown are the direct numerical results, using the
   numerical values of $C_{i,j}$ and $\kappa_i$, and the
   semi-analytical approximation obtained by solving numerically the
   discrete Poisson equation \eqref{geq} using the local
   equilibrium assumption $\kappa_i \simeq \rho_i / \beta^2$. In this
   and the following figures, each separate curve shows the
   stationary-state energy correlations $g_{i,j}$ for a given position $i$
   as a function of $j$.  (b) Rescaled correlations $L g_{i,j}$ in the
   absence of a density gradient, as a function of $j/L$, $i=(L+1)/2$
   and $i=(L+3)/4$, for the same parameter values as in (a) and
   different system sizes $L$. Onsite correlations $g_{i,i}$ are not
   shown. The exact result in the continuum limit is shown as a solid
   line.  }
\end{figure}

The simplest case with non-trivial correlations is to impose a
linear temperature gradient between two baths with temperatures $T_0
\neq T_{L+1}$, but with a flat concentration gradient, i.e.\ $\rho_0 =
\rho_{L+1} = \rho$ for all $i$, and without bias in the dynamics
($p=q$).  In this case, the density profile is flat throughout the
system, $\rho_i = \rho$ for all $i$.  The profile of mean energy is
\emph{linear}:
\begin{equation}
\mean{E_i} = \mean{E_0} + \frac{i}{L+1} \left( \mean{E_{L+1}} - \mean{E_0} \right).
\end{equation}
Identifying as usual $T_i \defeq \mean{E_i} / \rho_i$, we can conclude
that there is also a linear temperature profile under thse conditions;  this
is correctly obtained in simulations (not shown).

Figure~\ref{fig:temp-gradient-no-density-gradient} shows the
long-range part of the energy correlations, $\g{i}{j}$, in this case.
Numerical results are compared to a numerical solution of the
algebraic discrete Poisson equation~\eqref{eq:defn-g}. In order to
carry out this numerical solution, the source terms $\k{i}$ were
assumed to take their local equilibrium value $\rho/\beta_i^2$, as
described above. Despite this, we find very good agreement between the
numerical results obtained from direct simulation and the numerical
solution of the discrete diffusion equation. This holds everywhere,
including for the onsite contribution of $\g{i}{i}$.

\begin{figure}
 \includegraphics{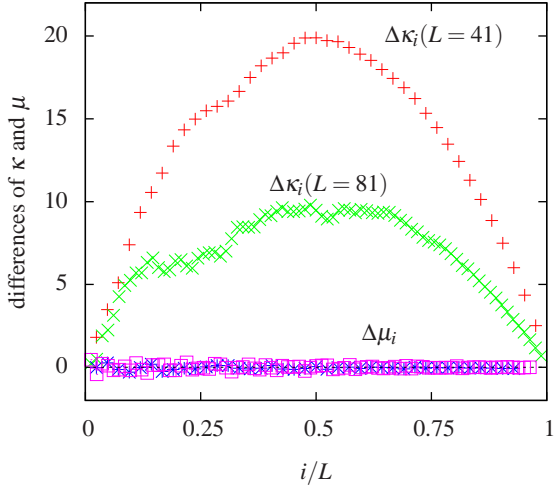}
 \caption{Difference $\Delta \kappa_i \defeq \kappa_i - \rho T_i^2$ of
   the value of $\kappa_i$ obtained from the direct numerical
   simulation and the analytical expression obtained with the local
   equilibrium assumption, for two values of system size $L$. The
   differences $\Delta \mu_i$ between the corresponding results for
   $\mu_i$ obtained by differenciation are also shown.}
 \label{fig:kappa-and-mu}
\end{figure}

Nonetheless, the agreement between the numerical and semi-analytical
results is affected by the fact that the local thermodynamic
equilibrium approximation is not strictly correct.  As discussed in
the introduction, the structure of the out-of-equilibrium measure is
an open problem.  However, here we can obtain an indication of the
error in the local thermodynamic equilibrium approximation by
comparing the value of $\kappa_i = \mean{e_i^2/(m_i+1)}$ obtained from
a direct numerical average to the analytical value $\rho_i /
\beta_i^2$ obtained from the local thermodynamic equilibrium
assumption.  This difference is shown in figure~\ref{fig:kappa-and-mu}
for two different values of system size $L$. We see that the marginal
distribution is not quite given by the local equilibrium
approximation, but that it gets closer as $L$ increases, in agreement
with the rigorous results of
\cite{RavishankarYoungLocalThermEqRandomHalvesJSP2007}.

For the structure of the correlations $\g{i}{j}$ away from the
diagonal terms where $i=j$, the important quantities are the
sources $\mu_i$, which are given by differences of the $\kappa_i$ as
in \eqref{eq:defn_mu}. The difference between the $\mu_i$ 
calculated by differentiating the numerically-obtained $\kappa_i$, and
those obtained by differentiating the local equilibrium expression, are
also shown in figure~\ref{fig:kappa-and-mu}. They are very close to $0$,
which is the reason for the excellent agreement between the numerical
and semi-analytical results for the correlations.

In fact, this case (absence of concentration gradient) is simple
enough to solve explicitly in the continuum limit. Taking $g(i,j)\to
G(x,y) \, \delta x \, \delta y$, equation \eqref{geq} can be rewritten in the
continuum limit as:
\begin{equation}
\frac{\partial G(x,y)}{\partial x^2} + \frac{\partial G(x,y)}{\partial
  y^2} =  4 c [T'(x)]^2 \, \delta(x-y),
\end{equation}
with boundary conditions $G(x=0,y)=G(x=L,y)=G(x,y=0)=G(x,y=L)=0$. The
solution of this equation is readily found to be
\begin{equation}\label{solution}
G(x,y)=\frac{2 c (\nabla T)^2}{L}
\left\{
\begin{array}{ll}
x(L-y), &\text{if\ } y>x \\
& \\
y(L-x), &\text{if\ } y<x
\end{array}
\right. .
\end{equation}
This result is similar to those of
refs.~\cite{NicolisMalekMansourOnsetSpatialCorrelationsPRA1984,
  BoonSuarezLongRangeCorrelationsLatticeGasPRE1996}, but with the
difference that the concentration $c$ now appears explicitly in the
result.

As was pointed out in
\cite{LinYoungCorrelationsRandomHalvesJSP2007}, the correlations for a
system of size $L$ decay as $1/L$ if the boundary conditions are fixed
(i.e. the values of the density and temperatures at the boundary are
the same for diferent values of the system size $L$). Figure~\ref{fig:scaling}
shows the rescaled correlations $L g_{i,j}$ for different system
sizes compared to \eqref{solution}.  In the figure we
have thus scaled space to the interval $[0,1]$ and rescaled the
correlations by multiplying them by $L$. The various curves indeed
converge to the limiting (continuum) form as $L\to \infty$.  
Note that the
apparent $1/L$ scaling arises from the $\delta x$ term in the passage to the
continuum limit: fixing the total system size and doubling the number of sites
corresponds to halving $\delta x$.

\subsection{Combined temperature gradient and concentration gradient}

\begin{figure}
\subfigure[$\rho_0 = 10$; $\rho_{L+1}=20$]
{
\includegraphics{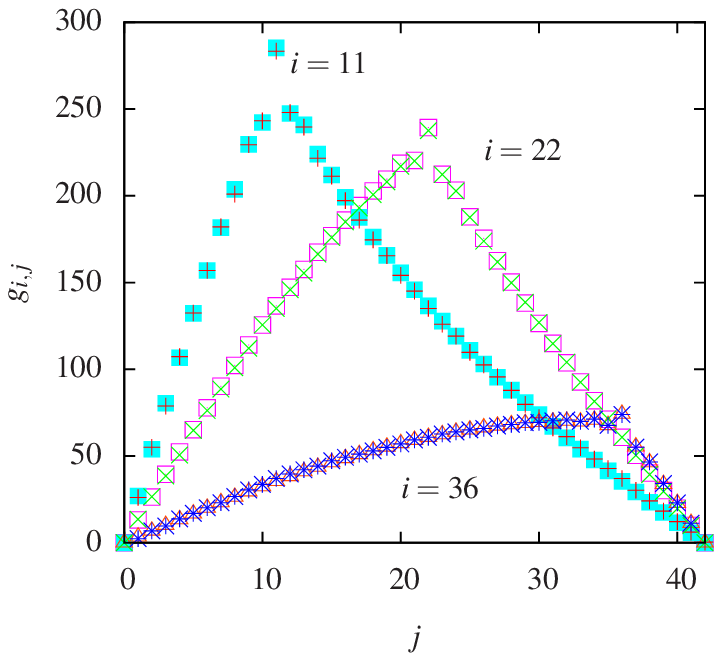}
}
\subfigure[$\rho_0 = 20$; $\rho_{L+1}=10$]
{
\includegraphics{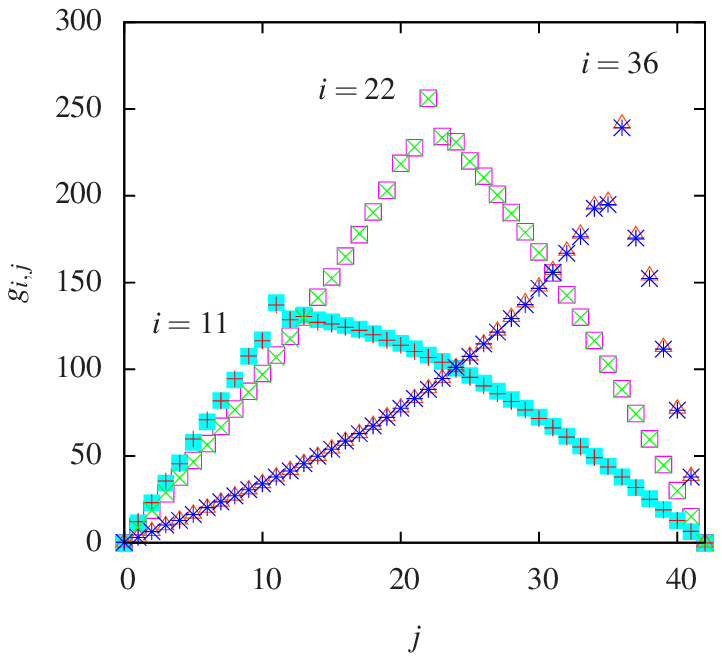}
}
 \caption{Long-range part of the correlations, $\g{i}{j}$, with imposed
temperature and density
   gradients. The parameters are $L=41$, $T_0 = 50$, and
   $T_{L+1}=10$. The bath densities are $\rho_0=10$ and
   $\rho_{L+1}=20$ in (a), and are reversed in (b).  The direct
   numerical and semi-analytical results again agree very well.  }
 \label{fig:both-gradients}
\end{figure}

We now consider the effect of imposing both energy and concentration
gradients, although still without bias in the motion ($p=q$).  The
profiles of $\rho_i$ and of $\mean{E_i} = \rho_i T_i$ are now both
linear, so that $T_i = \mean{E_i} / \rho_i$ is a ratio of two linear
functions, but is no longer itself linear.

\begin{figure}
 \includegraphics{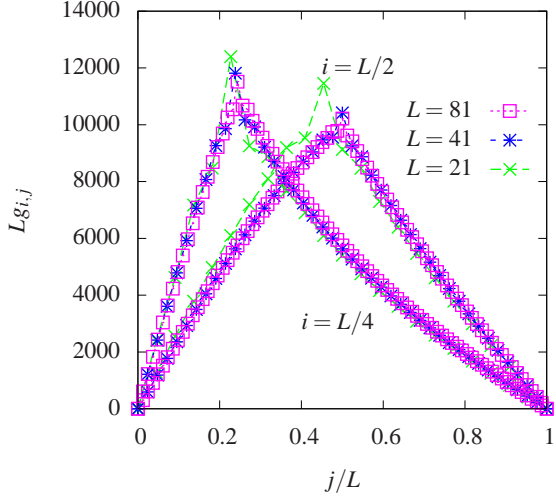}
 \caption{Scaling of correlations $L g_{i,j}$, obtained from the
   semi-analytical solution, for different system sizes $L$, with
   parameters as in figure~\ref{fig:both-gradients}(a), for
   $i=(L+1)/2$ and $i=(L+3)/4$.}
 \label{fig:scaling-with-gradients}
\end{figure}

Figure~\ref{fig:both-gradients} shows the numerical and
semi-analytical results in this case.  We find a skewing effect on the
correlations, which is again in excellent agreement with the numerical
results. The results are qualitatively very similar to those in
\cite{LinYoungCorrelationsRandomHalvesJSP2007}, despite the differences
in the nature of the models discussed in the introduction.
Figure~\ref{fig:scaling-with-gradients} shows the scaling of the
correlations (obtained from the semi-analytical results) with system
size. Again they converge to a continuum limit, corresponding to the
solution of the continuous diffusion equation with sources given by
\eqref{eq:continuum-source}.

\subsection{Effect of bias ($p \neq q$)}

Upon introducing a bias in the directionality of the walkers' jumps,
that is by putting $p \neq q$, we obtain mean density and energy
profiles which are no longer linear.  Rather, they are given by
\cite{LarraldeResendizStatsDiffusiveFlux2005}
\begin{eqnarray} 
\rho_i &= \rho_0 + \frac{1 - \alpha^i}{1 - \alpha^{L+1}} \, \left(\rho_{L+1}
 - \rho_0 \right),\\ 
\mean{E_i} &= \mean{E_0} + \frac{1 - \alpha^i}{1 -
   \alpha^{L+1}} \, \left(\mean{E_{L+1}} - \mean{E_0} \right),
\label{bias_profile}
\end{eqnarray}
where $\alpha \defeq p/q$ and the quantities $\rho_{L+1}$, $\rho_0$,
$\mean{E_{L+1}}$ and $\mean{E_0}$ are fixed by the boundary conditions.

\begin{figure}
\subfigure[]
{
 \includegraphics{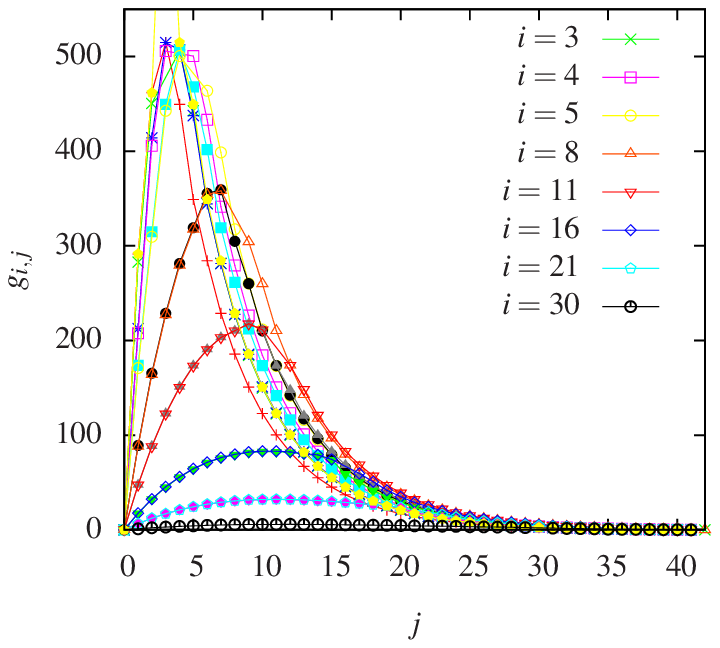}
  \label{fig:with-bias}
}
\subfigure[]
{
 \includegraphics{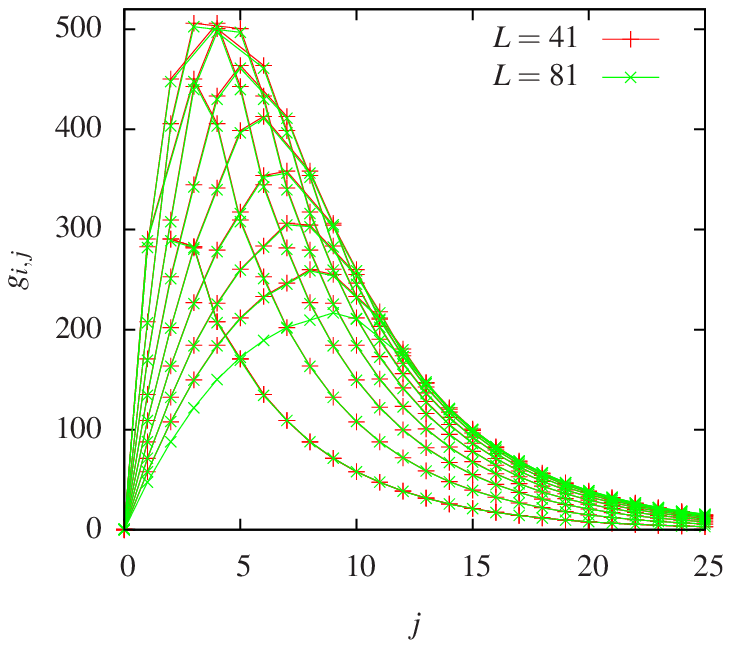}
  \label{fig:comparison-with-bias}
}
 \caption{(a) Comparison of numerical and semi-analytical correlations
   $g_{i,j}$ for system size $L=41$, with parameters $\rho=10$, $T_0 =
   50$, $T_{L+1}=5$, as a function of $j$, for $i=3$, $4$, $5$, $8$, $11$, $16$,
$21$ and $30$ from top to bottom.
There is a bias in the dynamics, with $p=0.35$
   and $q=0.4$. (b) Comparison of the correlations for $L=41$ and
   $L=81$, with the same parameters as in (a).  The numerical value of
   the correlations are the same for $i$ not too large. The lines are shown as a
guide for the eye}
\end{figure}

Figure~\ref{fig:with-bias} shows the comparison between the
numerically-obtained correlations and the semi-analytical solution,
for a situation with a flat density profile (i.e. $\rho_{L+1} =
\rho_0$), an imposed temperature gradient (which for constant density
means that $\mean{E_{L+1}} \neq \mean{E_0}$) and a bias to the
left. Again we find excellent agreement between the numerical results
and the semi-analytical results.  Note, however, that for values of $i$ around
30 and larger, the correlation function $g_{i,j}$ is very small for
all values of $j$ ($j\neq i$). Hence, the energies at sites
corresponding to large enough values of $i$ and $j$ are essentially
\emph{uncorrelated}. The reason for this is that, in the presence of a
bias, the source of the correlations given in
eq.\ \eqref{eq:continuum-source} decays exponentially as $i$ increases.
Thus, in this case, since the bias is to the left, we expect the
source of the correlations to be appreciable only up to distances of a
few times the decay length $\lambda\sim 1/|\ln(\alpha)|$ (see
eq.\ \eqref{bias_profile}) from the left boundary, and the effect of the
right boundary to become negligible if the size of the system
$L>>\lambda$. Indeed, figure~\ref{fig:comparison-with-bias} compares the
correlations for two different system sizes with the same
parameters. Clearly, the numerical values of the correlations are 
essentially independent of system size, and are negligible over
a large part of the system.


\begin{figure}
 \includegraphics{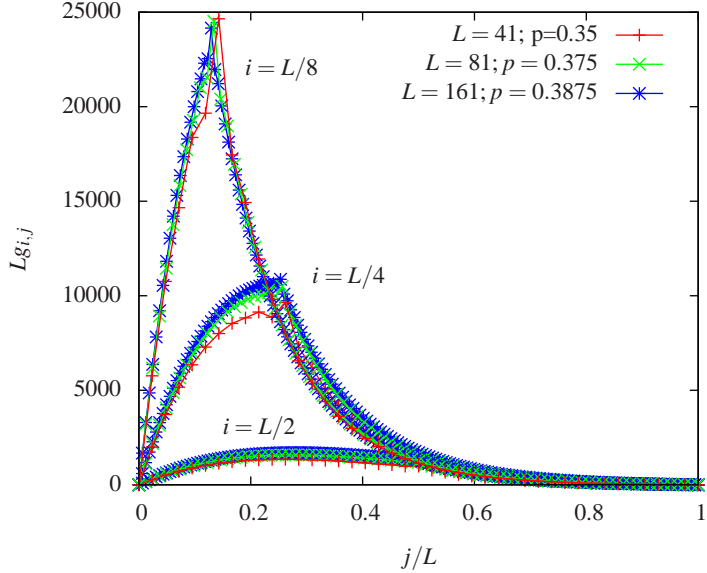}
 \caption{Rescaled semi-analytical correlations
   $L g_{i,j}$ for biased dynamics as a function of $j/L$, for system sizes
$L=41$, $L=81$ and $L=161$ and different $i$.
   In order to have a well-defined continuum limit, $q-p$ is halved when the
system size
is doubled, with $q=0.4$ fixed. Parameters are $\rho=10$, $T_0 =
   50$, $T_{L+1}=5$. 
}
  \label{fig:with-bias-scaling}
\end{figure}

Nonetheless, as was shown in section~\ref{sec:therm-forces}, it is possible to
obtain a well-defined continuum
limit for the energy correlations in the case of biased dynamics, provided the
amount of bias changes in the correct way: the bias $q-p$ must be halved
when the system size doubles, in addition to performing the same linear
rescaling of space as in the other cases.
Figure~\ref{fig:with-bias-scaling}  confirms convergence to the continuum
limit under these conditions.


\section{Conclusions}
By studying an extremely simple model of coupled transport of mass and
a second conserved quantity, which we called energy, we have shown
that the ubiquitous long-range correlations in this energy, whose
transport depends on the motion of the mass, are present in the
nonequilibrium stationary state, even though this quantity is completely
passive.  We were able to write down the equation describing the
long-range spatial correlations for the energy in the system, and
found that the structure of these correlations is remarkably similar
to those found in more realistic models, as well as to the results of
studies using fluctuating hydrodynamics. We thus conclude that the
origin of these long-range correlations is already present in this
simple model, and that a study of such models can go at least part of
the way to explaining and quantifying the origin and structure of
correlations in nonequilibrium systems.  Nevertheless, we hope to be
able to extend the methods and results to cases where the particle
motion is modified by the energy carried by the particles.

\section*{Acknowledgements}
DPS thanks the Erwin Schr\"odinger Institute and the Universit\'e
Libre de Bruxelles for hospitality, which enabled helpful discussions
with K.~Lin and L.~S.~Young, J.-P.~Boon, P.~Gaspard and T.~Gilbert. He
also thanks I. Santamar\'ia-Holek for useful comments.  Supercomputing
facilities were provided by DGSCA-UNAM.  Financial support from
PAPIIT-UNAM grant IN112307-3 and from the PROFIP programme of
DGAPA-UNAM are also acknowledged. We thank the anonymous referee for useful
comments, which improved the exposition of the paper.

\appendix
\section{Derivation of energy partitioning distribution and entropy at a site}

The distribution for the microcanonical partioning of energy among the particles at each
site can be calculated as follows. First, given that the particles at
each site are assumed to be in a state of microcanonical equilibrium,
we can determine the probability that $\lp{i}$ and $\lm{i}$ out of the $m_i$ particles at
site $i$ have combined energies $\sp{i}$ and $\sm{i}$ respectively, when the
total energy at the site is $e_i$, as the quotient of the number of states
consistent with these requirements over the total number of states
available to the system. These numbers are proportional to the
corresponding structure functions \cite{ReichlStatisticalPhysicsBook}, given by
\begin{equation}
\w(E,N,V) \defeq \int \delta \left( H_N({\mathbf{p}},{\mathbf{q}})-E \right) \,
{\mathrm d} {\mathbf{p}} \, {\mathrm d}{\mathbf{q}},
\end{equation}
where $H_N({\mathbf{p}},{\mathbf{q}})$ is the $N$-particle Hamiltonian
describing the dynamics at each site, and the integration is carried
out over the phase space of the $N$ particles. If we assume that the
particles are an ideal gas, then
\begin{equation}
H_N({\mathbf{p}},{\mathbf{q}})=\sum\limits_{j=1}^N\frac{|p_i|^2}{2m},
\end{equation}
where $m$ is the mass of the particles, and the required probability
is given by
\begin{eqnarray}\label{condprob}
\fl
\condprob{\sp{i},\sm{i}} {\lp{i},\lm{i},m_i, e_i} = \\
\frac{\w(\sp{i},\lp{i},V) \, \w(\sm{i},\lm{i},V) \, \w(e_i-\sp{i}-\sm{i},m_i-\lp{i}-\lm{i},V)}{\w(e_i,m_i,V)}.
\end{eqnarray}
For a $d$-dimensional ideal gas, we have
\begin{equation}
\w(E,N,V)=\frac{(2\pi m)^{Nd/2} E^{Nd/2 - 1} V^N}{\Gamma(Nd/2)},
\end{equation}
where $V$ is the volume accesible to the particles at each site, which
we take as unity, and $\Gamma(\cdot)$ is the gamma function. Using this expression in \eqref{condprob} yields
\begin{eqnarray}\label{condprob2}
\fl
\condprob{\sp{i},\sm{i}} {\lp{i},\lm{i},m_i, e_i} = \\
\frac{\Gamma(m_id/2)}{\Gamma(\lp{i}d/2)\Gamma(\lm{i}d/2)\Gamma([m_i-\lp{i}-\lm{i}]d/2)} \times \\
(\sp{i})^{\lp{i}d/2-1} (\sm{i})^{\lm{i}d/2-1}
\frac{(e_i-\sp{i}-\sm{i})^{[m_i-\lp{i}-\lm{i}]d/2-1}}{e_i^{m_id/2 -1}},
\end{eqnarray}
which, when $d=2$, simplifies somewhat, giving
\begin{eqnarray}
\fl
\condprob{\sp{i},\sm{i}} {\lp{i},\lm{i},m_i, e_i} = \\
\fl
\frac{\Gamma(m_i)}{\Gamma(\lp{i})\Gamma(\lm{i}) \Gamma(m_i-\lp{i}-\lm{i})} 
(\sp{i})^{\lp{i}-1}  (\sm{i})^{\lm{i}-1}
\frac{(e_i-\sp{i}-\sm{i})^{m_i-\lp{i}-\lm{i}-1}}{e_i^{m_i-1}}.
\end{eqnarray}

Similarly, the classical entropy of the 2D gas at each site
is given by \cite{CallenThermodynamicsBook1985}
\begin{equation}
S(e_i,m_i,V)= \ln[\w(e_i,m_i,V)/m_i!] + m_i \sigma_i \sim  m_i \ln[V e_i
  /m_i^2] + m_i s_i,
\end{equation}
 taking the Boltzmann constant $k_B=1$,
where $s_i$ and $\sigma_i$ are constants (i.e.\ they are independent of
$e_i$, $m_i$ and $V$) which can vary from site to site.

\section*{References}

 \bibliographystyle{naturemag}

\bibliography{long-range}

\end{document}